\documentclass[aps,prl,preprint,superscriptaddress,showpacs]{revtex4}
\usepackage{graphics}

\def \b {\begin{eqnarray}}
\def \el#1 {\label{#1} \e}
\def \e {\end{eqnarray}}
\def \ds {\displaystyle }

\newcommand{\ave}[1]{\langle {#1} \rangle}

\begin{document}

\preprint{version 2}

\title{Dynamics of Air-Fluidized Granular System Measured by the Modulated Gradient Spin-echo}

\author{Janez Stepi\v{s}nik}
\email[]{Janez.Stepisnik@fiz.uni-lj.si}
\affiliation{University of Ljubljana, Faculty of Mathematics and Physics, Physics Department, Jadranska 19, 1000 Ljubljana, Slovenia}
\altaffiliation{Josef Stefan Institute, Jamova 39, 1000 Ljubljana, Slovenia}

\author{Samo Lasi\v c}
\affiliation{University of Ljubljana, Faculty of Mathematics and Physics, Physics Department, Jadranska 19, 1000 Ljubljana, Slovenia}

\author{Igor Ser\v sa}
\affiliation{Josef Stefan Institute, Jamova 39, 1000 Ljubljana, Slovenia}

\author{Ale\v{s} Mohori\v{c}}
\affiliation{University of Ljubljana, Faculty of Mathematics and Physics, Physics Department, Jadranska 19, 1000 Ljubljana, Slovenia}

\author{Gorazd Planin\v si\v c}
\affiliation{University of Ljubljana, Faculty of Mathematics and Physics, Physics Department, Jadranska 19, 1000 Ljubljana, Slovenia}

\date{\today}

\begin{abstract}
The power spectrum of displacement fluctuation of beads in the air-fluidized granular system is measured by a novel NMR technique of modulated gradient spin-echo. The results of measurement together with the related spectrum of the velocity fluctuation autocorrelation function fit well to an empiric formula based on to the model of bead caging between nearest neighbours; the cage breaks up after a few collisions~\cite{Menon1}. The fit yields the characteristic collision time, the size of bead caging and the diffusion-like constant for different degrees of system fluidization. The resulting mean squared displacement increases proportionally to the second power of time in the short-time ballistic regime and increases linearly  with time in the long-time diffusion regime as already confirmed by other experiments and simulations.
\end{abstract}


\maketitle

\section{Introduction}

Sand dunes, grain silos, building materials, catalytic beds, filtration towers, riverbeds, snowfields, and many foods are granular systems consisting of large number of randomly arranged macroscopic grains. Despite their apparent simplicity granular materials exhibit a host of unusual behaviours, whose unravelling more often than not appears to challenge existing wisdom of science~\cite{Jaeger,deGennesgrain}.

Fluidized granular bed is a system of randomly arranged, macroscopic grains in which the driving force of motion is container shaking or gas flow through the granular system. Although, these systems are of tremendous technological importance in catalysis of gas-phase reactions, transport of powders, combustion of ores, and several other industrial processes, we do not have sufficient understanding of the fluid state of granular medium that is analogous to macroscopic properties of liquids. Two particularly important aspects contribute to the unique properties of granular materials: thermodynamics plays no role, and interactions between the grains are dissipative, because of static friction and inelasticity of collisions. Several theoretical efforts start towards building granular fluid mechanics by considering the medium as a dense, inelastic gas with the temperature defined by induced local velocity fluctuations~\cite{Bagnold,Jenkins}. The autocorrelation function of velocity fluctuation is the basis of many thermodynamic and hydrodynamic models which aim to provide a statistical description of a granular system in terms of a single particle dynamic. In the simulation of hard-sphere fluid, Alder and Wainwright~\cite{Alder1} found a strong dependence of the velocity autocorrelation function (VAF) on the system density. Only for very low particle densities, it decays exponentially with the Enskog correlation time~\cite{Enskog}, while at higher densities, a negative long-time tail appears as the result of caging by adjacent spheres.

Although, the experimental techniques used to study the motion of granular systems span a wide range of approaches and sophistication, very few attempts were made to finger into the details of grain motion. To the best of our knowledge, only the tracking of bead motion by the positron emission~\cite{Wildman2} and  by the CCD camera~\cite{Utter} were able to glimpse to the VAF of a fluidized granular bed.

The NMR gradient spin-echo is a tool that yields not only macroscopic but also microscopic dynamic variables due to the relation between spin-echo attenuation  and VAF of spin bearing particles~\cite{deGennes,moj81,moj202}. However, this potential of spin-echo has been only partially exploited~\cite{mojcall3,moj001,Codd,Topgaard,Parsons}, particularly, when used for the study of granular motion~\cite{Seymourgrain,Seymourgrain2}. 

In this letter, we report on the first application of modulated gradient spin-echo (MGSE) to measure autocorrelation spectra of motion in a fluidized granular bed. MGSE is a method, in which a repetitive train of radiofrequency (RF) pulses with interspersed magnetic field gradient pulses or a gradient waveform periodically modulates the spin phase, so that the spin-echo attenuation is proportional to the power spectrum of the spin displacement fluctuation (PSDF) or the spectrum of VAF. The frequency range of the measurement is determined by the rate of spin phase modulation, which can be between a few Hz to about a few $10$ kHz at the present state of NMR hardware. This covers the expected range of fluidized bead motion~\cite{Menon1}. Generally, the measurement of the fluctuation autocorrelation by the MGSE method can be considered as a low-frequency complement to the non-elastic neutron scattering method that covers the range above GHz frequency.
\begin{figure}[ht]
\centering \scalebox{0.9}{\includegraphics{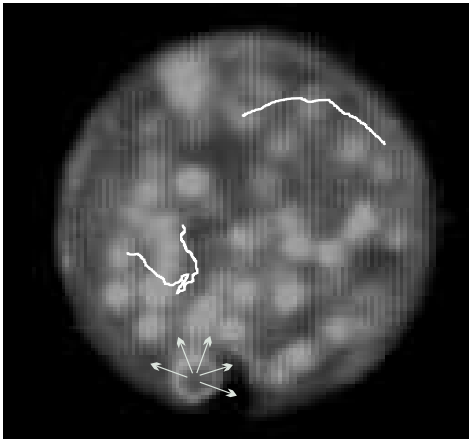}} \caption{ The high-speed camera snap-shot of grains moving in the air fluidized system. The curves show the paths of two representative grains during $20$ ms long MGSE sequence. The arrows at the bottom indicate the air flow inlet through the perforated hose.} \label{fig.1}
\end{figure}

When using NMR spin-echo to detect the translational motion of a spin-bearing particle, we observe the cumulative effect of small perturbations of spin precession frequency due to the displacements of the spin-bearing particle in the magnetic field gradient. The random collisions between grains in the fluidized granular system shift a bead from the path of mean motion driven, for example, by the flow of air. As the phase of the spin-echo signal depends on the displacements along the mean path of motion, the signal attenuation depends on the PSDF,  $ I_z(\omega )$ , as~\cite{moj81}
 \b
\beta(\tau)=\frac{\gamma^2 }{2\pi}\int_{-\infty}^\infty I_z(\omega )\vert {\bf G}( \omega ,\tau )\vert^2\,d\omega,
\e 
whenever the displacement fluctuations from the mean path are short compared to the phase shift grating created by the applied gradient~\cite{moj993}. The spectrum of the effective gradient ${\bf G}_{eff}(t)$ is defined as~\cite{moj81}
\b
{\bf G}(\omega,\tau  )=\int_0^{\tau}\,{\bf G}_{eff}(t) e^{-i\omega t}\,dt,
\e
where $\tau$ is the time of spin-echo measured from the first excitation pulse. The effective gradient changes the sign upon every application of the $\pi$-RF pulse. According to the Wiener-Khintchine, the PSDF is related to the displacement autocorrelation function~\cite{Kubo2} 
\b
I_z(\omega ) =\frac{1 }{\pi} \int_{-\infty}^{\infty}
\ave{
\Delta z(t)\, \Delta z(0) } e^{\displaystyle{-i\omega
t}} dt,
\el{DAF}
where $\Delta z(t)=z(t)-\ave{z(t)}$ denotes the displacement fluctuation along the direction of applied gradient with $\ave{z(t)}$ being the mean spin location  at time t. The PSDF is related to the power spectrum of the velocity fluctuation or the spectrum of the VAF
\b 
D(\omega)&=&I_z(\omega)\,\omega^2,
\el{avto}
and to the mean squared displacement of grain fluctuation
\b 
\ave{[\Delta z(t)-\Delta z(0)]^2}&=&\frac{4 }{\pi}\int_{0}^\infty  I_z(\omega)( 1-\cos(\omega t))d\omega.
 \el{MSD}
\begin{figure}[ht]
\centering \scalebox{0.8}{\includegraphics{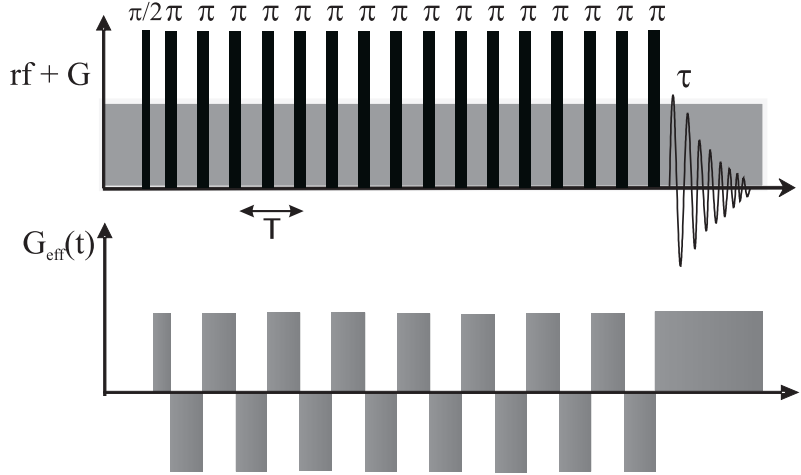}} \caption{ RF-gradient sequence that yields the spectrum of effective gradient with the dominant peak at the frequency $\omega= 2\pi/T$.} \label{fig.2}
\end{figure}

The fluctuating part of displacement cannot be distinguished from the mean displacement with the use of the conventional spin-echo with two gradient pulses (PGSE) ~\cite{Savelsberg,Mair}, where the resulting propagator includes the average of a grain flow and grain displacement fluctuations. The flow compensating four gradient pulse variant of PGSE ~\cite{Seymourgrain} is unsuitable for the PSDF sampling, because its gradient spectrum covers a broad-frequency range~\cite{Callaghan}. However, a RF-gradient sequence, that causes a cyclic spin phase modulation, yields a gradient spectrum with sharp peaks. After $N$ modulation cycles of the period $T$, the resulting spin-echo attenuation is
 \b
\beta(N T)&=&2\gamma^2\,N T\sum_{n}I_z(n\omega_m )\vert {\bf g}(n\omega_m)\vert^2.
\el{att}
Here $I_z(n\omega_m)$ are values of the PSDF at the peaks of the gradient spectrum at $\omega= n\frac{2\pi}{T}$ with $n=0, \pm 1, \pm 2, \pm 3...$. The peak amplitudes are ${\bf g}(\omega)=\frac{1}{T}\int_0^T\,{\bf G}_{eff}(t) e^{-i\omega t}\,dt$, while the peak widths are $2\pi/NT$. A sequence with a single dominant peak in the gradient spectrum~\cite{mojcall} can be designed with a proper phase cycling.  The CPMG train of $\pi$ RF pulses~\cite{Meiboom}  applied to spins in a constant magnetic field gradient with magnitude $G$ forms  dominant peaks at $\omega_m=\pm 2\pi/T$,  whenever the train of $\pi$-RF pulses, which are repeated at the  $T/2$ intervals, is applied a quarter of the period after the excitation $\pi/2$-RF pulse, as shown in Fig.\ref{fig.2}. The resulting spin-echo attenuation at $\tau=NT$ 
 \b
\beta(NT, \omega_m)&=& \frac{8\gamma^2{ G}^2\,NT}{\pi^2}I_z(\omega_m),
\el{MGSE}
can be used to sample the PSDF by varying the modulation period $T$ . 

 \section{Experimental procedure}

The experiment was carried out on a TecMag NMR spectrometer with a $2.35$ T horizontal bore superconductive magnet. The spectrometer was equipped with  micro-imaging accessories and with reversed Helmholtz gradient coils with 0.25 T/m peak magnetic field gradient.
 
The described MGSE method was applied to a granular system made up of 100 pharmaceutical 3-mm diameter, oil-filled, hard plastic spherical beads with the restitution coefficient 0.85 placed in a cylindrical chamber. The container, build from a piece of a plastic syringe tube of 26 mm length and 23.5 mm diameter, was placed inside the RF coil with the cylinder axis being horizontal. A perforated hose of 5 mm outer diameter, attached to the bottom inner wall of the container, is used as a diffuser of air up-flow as shown in Fig.\ref{fig.1}. Several holes of 0.5 mm diameter are drilled uniformly across the surface of the container to serve as the air outlets. The degree of fluidization was regulated by the air pressure in the hose and monitored by a high speed camera (600 fps) through the transparent walls of the cylindrical container. At applied air pressures, we observed that initial tendency of uplifted beads, to circulate around the container walls, breaks down by collisions and dispersed air flow leading to a random motion of beads.  The system was considered as a fully fluidized because a distinct bead clustering or bubbling was not observed~\cite{Volfson} . Fig.\ref{fig.1} shows two paths of representative single grains at the air pressure of $0.5$ bar during 10 steps that correspond to 20 ms. The paths of beads in the bulk are short enough that the collisions with the container walls are seldom during the measuring interval.    

The PSDF of fluidized system was measured in the static gradient field of $0.0625$ T/m in the vertical direction, perpendicular to the axis of cylindrical container. The axis was parallel to the static magnetic field. The amplitudes of spin-echo were recorded at different modulation periods, $T$, while keeping the acquisition time constant, $t=NT=20$ ms. The co-addition of echoes, taken at different steps of RF phase  cycle of CPMG train,  is used to remove the echoes of spurious coherence path-ways as well as a residual phase shift of the grain flow. Thus, the applied method was able to detect only the displacement fluctuation caused by the collisions and the air flow. 
 
 \section{Results and discussion}
 \begin{figure}[ht]
 \centering \scalebox{0.8}{\includegraphics{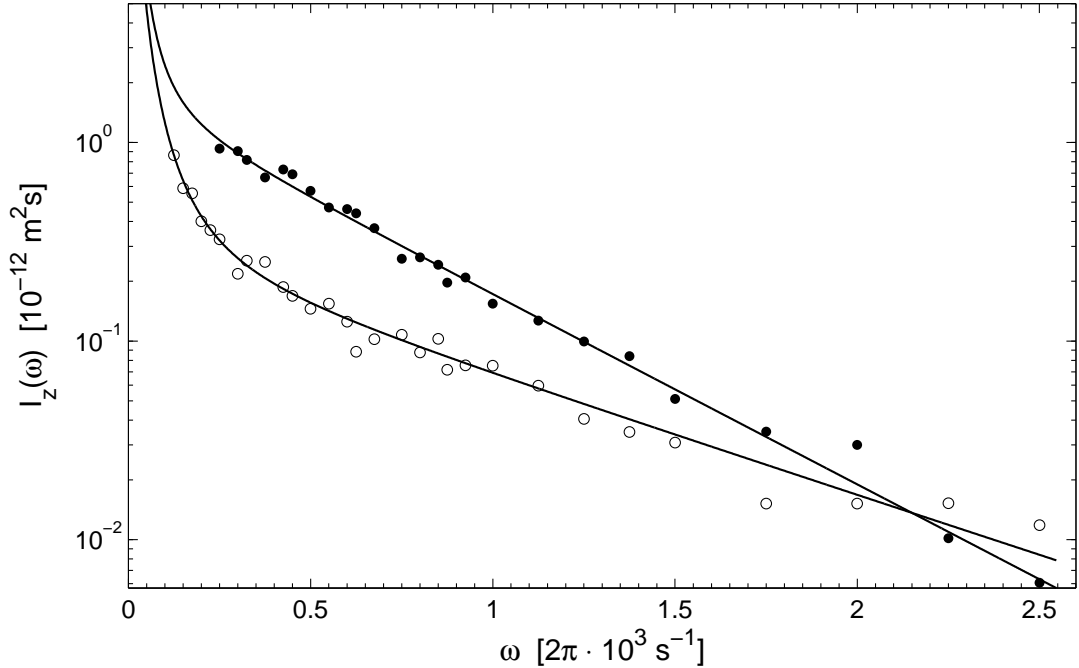}} \caption{The displacement fluctuation spectra of oil-filled beads fluidized by the air-flow at pressures of $0.25$ bar (empty circles) and $0.5$ bar(full circles) . At higher frequencies, the experimental points indicate clear exponential decay. At the low pressure, there is a very distinctive $\frac{1}{\omega^2}$-dependence at low modulation frequencies. Curves display the best fit by the empiric formula (Eq.\ref{empiric}).} \label{fig.3}
 \end{figure}
\begin{figure}[ht]
   \centering \scalebox{0.8}{\includegraphics{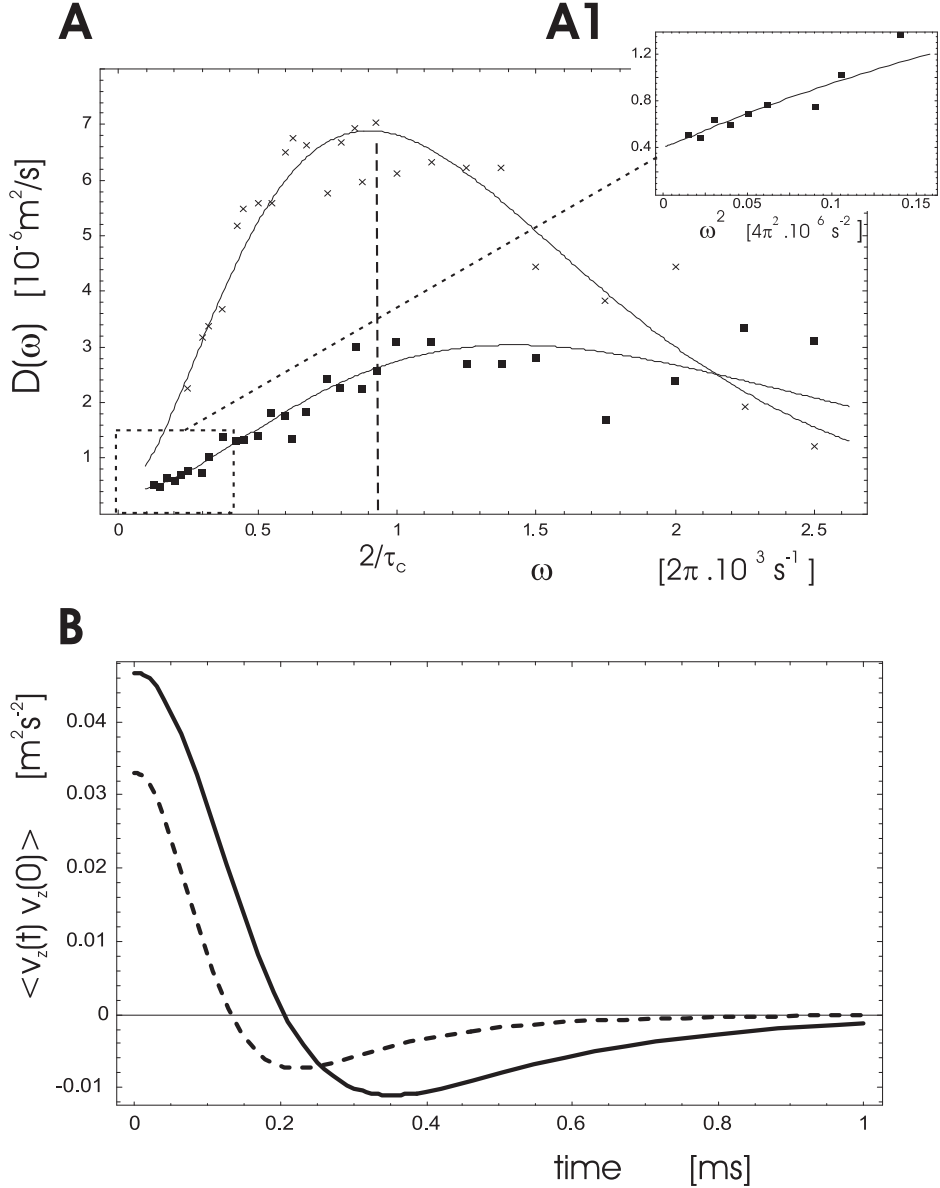}} \caption{ A.) The power spectrum of grain velocity fluctuation in the system of oil-filled beads fluidized by the air-flow at 0.25 bar (squares) and 0.5 bar (crosses). A1.) The intersection of line with the ordinate in $D(\omega)$ vs. $\omega^2$-plot gives the diffusion-like constant. B.) The Fourier transform of $D(\omega)$ gives the VAF of beads for air flow at the pressure $0.25$ bar (dashed) and $0.5$ bar.}\label{fig.4}
 \end{figure}

The MGSE measurement gives $I_z(\omega)$ of beads in the  air-fluidized system at different gas pressures, as shown in Fig.\ref{fig.3}. The spectra exhibit a clear exponential decay at frequencies above $400$ Hz, which is more distinctive for stronger air-flow at 0.5 bar. Evidently, faster grain motion increases the spin-echo attenuation at higher modulation frequencies magnifying the high-frequency range of measurement, but hindering the examination below $300$ Hz, because of a weak signal. At the reduced air-pressure of 0.25 bar, the MGSE measurement can trace $I_z(\omega)$ below $400$ Hz, yielding the frequency dependence that passes from an exponential at high frequencies into a $1/\omega^2$ low-frequency dependence.  The power spectrum of VAF, $D(\omega)$ in Fig.\ref{fig.4}A, calculated according to Eq.\ref{avto}, is similar to the power spectrum of over-damped thermal harmonic oscillators~\cite{Wang45}. The form of the spectrum is not Lorentzian as one would expect by assuming the Enskog exponential decay of the velocity autocorrelation function. Interestingly, the positron emission measurements give a similar spectral lobe for the vibrofluidized granular bed~\cite{Wildman2}, which is explained as a bead caging within the experimental cell. Better signal to noise ratio of MGSE measurements enables a detailed analysis of $I_z(\omega)$ and related $D(\omega)$. In Fig.\ref{fig.4}A1, $D(\omega)$ shows a clear $\omega^2$-dependence at  low frequencies, which is typical for  restricted diffusion. In restricted diffusion case, the slope of $D(\omega)$ vs. $\omega^2$ plot provides information about the size of  spin confinement, while the intersection with the ordinate, $D(0)$, gives the inter-pore diffusion rate in a system with inter-pore channels or permeable walls. Thus, a small value of $D(0)$ in Fig.\ref{fig.4}A1 could be interpreted by the  model of bead ballistic motion  between successive collisions, where many collisions are required to break out the caging formed by adjacent beads. The exponential frequency dependence of $I_z(\omega)$ in the high-frequency range, as observed in our experiments, is  different from the Lorentzian spectrum for diffusion in porous media~\cite{moj001,moj201}. However, the length of ballistic grain motion depends on the distribution of adjacent beads, which might be responsible for the observed form of $I_z(\omega)$ at high frequencies. By taking into account the experimentally observed character of the spectrum, an empiric formula 
 \b
I_z(\omega)=\frac{D+\ave{\xi^2}\tau_c\omega^2}{\omega^2}\,e^{\ds{-\tau_c\omega}}
 \el{empiric}
is introduced, which gives  the low-frequency dependence similar to the restricted diffusion ~\cite{moj201}, and the exponential decay at high frequencies.  
Here $\sqrt{\ave{\xi^2}}$ is the size  of the bead caging, $\tau_c$  is the characteristic time of ballistic motion and  $D$ is the diffusion-like constant describing hopping between cages. 
The mean squared displacement, calculated from this formula according to Eq.\ref{MSD}, is $\ave{\Delta z^2(t)} \approx 4\ave{\xi^2}t^2 /\pi\tau_c^2$ in the short-time ballistic regime and  is $\ave{\Delta z^2(t)}\approx 2 D t $ in the long-time diffusion regime. The optical measurements~\cite{Menon2} and the simulations~\cite{Wojcik} give a similar time-dependence.

The curves in Fig.\ref{fig.3} correspond to the best fits of the empiric formula to the experimental data. For the air-flow at $0.25$ bar, the fitting parameters are: $\ave{\xi^2}=1.2\times 10^{-9}$  $\textstyle{\rm m^2}$, $\tau_c=0.22$ ms and $D=0.47\times 10^{-6}$  $\textstyle{\rm m^2/s}$, with the relative error of  $5\%$. The air-flow at $0.5$ bar gives a very clear exponential dependence with the fitting parameters: $\ave{\xi^2}=4.4\times 10^{-9}$ $\textstyle{\rm m^2}$, and $\tau_c=0.36$ ms, but less exact $D=0.6\times 10^{-6}$ $\textstyle{\rm m^2/s}$, because of the attenuation cut-off in the low-frequencies range. The characteristic time of ballistic motion related to the bead collision frequency can be obtained from the position of $D(\omega)$-maximum  at $\omega=2/\tau_c$.
 
The Fourier transform of $D(\omega)$, modelled with the above fitting parameters, gives the VAF of fluctuation, $\ave{\Delta v_z(t)\,\Delta v_z(0)}$, as shown in Fig.\ref{fig.4}B. Its intersection with the ordinate provides the mean squared of the velocity fluctuation, which is by definition proportional to the temperature of the fluidized system: $T\approx\ave{v_z^2(0)}$.  It demonstrates an increase of the system temperature by about $45\%$ as the pressure, driving the air-flow, increases from $0.25$ bar to $0.5$ bar.

\section{Conclusion}
The new MGSE method provides the PSDF and the spectrum of VAF of beads in the air fluidized granular system. The data fit well to the empiric formula that corresponds to the model of bead caging within the space of adjacent colliding beads, which breaks up after many collisions. In references~\cite{Wildman1,Wildman2}, the positron emission measurement result in a similar spectra shape of VAF for the grain motion in a vibrofluidized granular system. Similar VAF, as shown in Fig.\ref{fig.4}B, was measured in the shear flow of a granular system by using a CCD camera~\cite{Utter}, where a break down of the exponential  decay at high particle densities was reported. In the statistical sense, the observed exponential form of the PSDF could be related to the observed exponential distribution of forces in a sheared granular bed as reported in Ref.~\cite{Jaeger}.

\begin{acknowledgments}
We are grateful to the Slovenian Ministry for High Education, Science and Technology for financial support.  One of us (JS) recalls a stimulating discussion with Dr. E. Fokushima, New Mexico Resonance Group, concerning alternatives to the measurement of grain dynamics, and who provided the sample for our measurements.
\end{acknowledgments}


 \end{document}